\documentclass[11pt]{article}

\usepackage{amsmath,amsfonts}

\makeatletter

\@addtoreset{equation}{section}

\newtheorem{theorem}{Theorem}
\newtheorem{lemma}{Lemma}
\newtheorem{remark}{Remark}

\def\J{{\bf J}}
\def\W{{\bf W}}
\def\M{{\bf M}}
\def\x{{\bf x}}

\def\m{{\bf m}}
\def\u{{\bf u}}
\def\e{{\varepsilon}}
\def\N{{\cal N}}
\def\P{\hbox{ Prob}}

\def\sumd{\displaystyle\sum}
\def\intd{\displaystyle\int}
\def\prodd{\displaystyle\prod}

\begin{document}

\title{Stability of the dynamics of an asymmetric neural network}
\author{J. F. Feng
$^\dagger$ \qquad M. Shcherbina $^\ddagger$ \qquad B. Tirozzi $^*$
\\
$^\dagger$
Department of  Computer Science and Mathematics\\
 Warwick
University, Coventry CV4  7AL, UK
\\
$^\ddagger$Institute for Low Temperature Physics, Ukr. Ac. Sci., 47
Lenin Ave.\\
 Kharkov, Ukraine \\
$^*$Department of Physics, Rome University 'La Sapienza', 00185
Roma, Italy
 }
\date{}
\maketitle
\begin{abstract}
We study the stability of the dynamics of a network of $n$ formal
neurons interacting through an asymmetric matrix with independent
random Gaussian elements of the type introduced by Rajan and
Abbott ((\cite{AR}). The neurons are represented by the values of
their electric potentials $x_i, i=1,\dots,n$. Using the approach
developed in a previous paper by us (\cite{FST1}) we obtain
sufficient conditions for diverging synchronized behavior and for
stability.
\end{abstract}

\section{Introduction}

The dynamic of a system of neurons described by their electric
potentials $x_i(t), i=1,\dots,n$ interacting linearly through a
random matrix has been extensively studied in the past literature
and received increased attention in the last times, see for
example (\cite{[CN]}, \cite{[91]}, \cite{[92]}), \cite{[pre]}).
The first statement about the stability of the solution of the
system

\begin{equation}
\x  '=-\kappa\x+{\J}'\x,
 \label{dynintr}\end{equation}

was enunciated by May (\cite{[May]}). Here  ${\J}'$ is a real
symmetric $n\times n$ matrix with independent gaussian elements
and $E {\J}'_{ij}=0$, $E {\J}^{'2}_{ij}=1/n$. The conjecture was
that if $ \kappa > \lambda_{\text{max}}$, where $
\lambda_{\text{max}}$ is the maximum eigenvalue of ${\J}'$, then
the solutions of the system (\ref{dynintr}) were stable. This
conjecture has been proved by us many years later in the paper
(\cite{FST1}) where the self-averaging property of the system have
been used. In particular we introduced the random counting measure

\begin{equation}
\N_n(\lambda,t)=n^{-1}\sharp\{x_i(t)\le\lambda\}=n^{-1}\sum_{i=1}^n\theta(\lambda-x_i(t)),
 \label{Nintro}\end{equation}

where $\theta(x)$ is the standard Heaviside function. This
function $\N_n$ counts the fraction of the electric potentials
$x_i(t)$ which are less than a given threshold $\lambda$. The
self-averaging property of $\N_n$ means that $\N_n(t) \to E
\N_n(t)$ in the $L^2$ norm with respect to the probability measure
of the gaussian matrix, $E$ being the expectation with respect to
the probability of all the random entries of the matrix ${\J}'$.
In (\cite{FST1}) we were able to proof this property, so the
random measure becomes asymptotically a gaussian distribution
function with mean value $a(t)$ and dispersion $\sigma(t)$:

\begin{equation}
\lim_{n\to\infty}E\{\N_n(\lambda,t)\}=\int_{-\infty}^\lambda
dx\frac{e^{-(x-a(t))^2/2\sigma(t)}}{\sqrt{2\pi\sigma(t)}}.
 \label{erfintr}\end{equation}

The results of the calculations were that $a(t)=e^{-\kappa t}$ and
$\sigma(t)=e^{-2\kappa t}J_0(wt)$ where $J_0$ is the Bessel
function of zero order. From the asymptotic behavior of $J_0$ we
get that if $ \kappa > w$ $\sigma(t)$ goes to zero and we get a
stable solution. It is a quite remarkable coincidence that this
nice result depends on the self-averaging property of the system,
this shows the real power of such property if one reminds all the
rigorous properties which have been possible to show in the field
of Statistical Mechanics of disordered systems. In this paper we
look for analogous results when a different matrix ${\J}'$ is
considered. The elements of ${\J}'$ are still independent but each
row of the matrix have mean values depending on the column index:

\begin{equation}
E\{J_{ij}'\}=a\cdot\left\{\begin{array}{ll}
\mu_I/n^{1/2},&j=1,\dots [fn],\\
\mu_E/n^{1/2},& j=[fn]+1,\dots,n.\end{array} \right \}
\end{equation}

The first $[fn]$ columns represent inhibitory interaction ($\mu_I
<0$) while the other $n-[fn]$ are excitatory interaction ($\mu_E
>0$), thus each neuron $i$ receives $[fn]$ inhibitory inputs of the
same type from the other neurons and $n-[fn]$ excitatory inputs
form the other neurons, the excitation and the inhibition do not
depending on the particular neuron $i$. With this choice the
matrix ${\J}'$ is asymmetric and the variances of the matrix
elements of ${\J}'$ also follow the same choice:

\begin{equation} E\{(J_{ij}'-E\{J_{ij}'\})^2\}=n^{-1}\sigma_j=\left\{\begin{array}{ll}
\sigma_I/n,&j=1,\dots [fn],\\
\sigma_E/n,&j=[fn]+1,\dots,n.\end{array} \right.
\end{equation}

Thus we look at the same property as before in the case of this
new matrix which includes inhibitory and excitatory inputs which
is nearer to realistic neural interactions. In order to understand
better the new kind of stability properties that we obtain let us
introduce some more definitions. Let $\m=(m_1,\dots,m_n)$ be the
vector defined by

\begin{equation} m_i=\left\{\begin{array}{ll}
\mu_I/n^{1/2},&j=1,\dots [fn]\\
\mu_E/n^{1/2}&,j=[fn]+1,\dots,n\end{array} \right.
  \label{mintro}
  \end{equation}

and $\M$ be the matrix with all rows equal to $\m$. Then in the
paper, following the ideas of ((\cite{AR}), we introduce the
decomposition

\begin{equation}\label{J+M1}
   \J'=\J+a\M,
\end{equation}

in order to have all the eigenvalues included in the circle with
radius one in the complex plane. The two theorems shown in this
paper describe in detail the stability and asymptotic properties
of the dynamic associated to the matrix $\J'$ and in particular
these properties are very sensible to the choice of the initial
conditions for the $x_i(t)$. We give here some hint since the
complete definitions will be given in the next section. So suppose
that the initial conditions can be written in the following way:

\begin{equation}\label{in_cint}
x_i(0)=c_i+\xi_i,
\end{equation}

where $\{\xi_i\}$ are independent random variables with
distributions $\{\nu_i\}$, satisfying the conditions

\begin{equation}\label{x0int}
    \quad E\{\xi_i\}=0,\quad E\{(\xi_i)^2\}=\sigma^{(0)}_i,\quad E\{(\xi_i)^8\}\le C.
\end{equation}
and the initial constants depend on the neuron in the same way as
the
\begin{equation}\label{cint}
    c_i=\left\{\begin{array}{ll}
c_I,&i=1,\dots [fn],\\
c_E,& otherwise,\end{array}\right.\end{equation}

In this situation the theorems proved in the paper establish that
the contribution to the dynamic of the matrix $\J$ is stable if
$\kappa >\sigma_*=f\sigma_I+(1-f)\sigma_E$. Remark that since the
matrix $\J$ is asymmetric the stability of the dynamic does not
depend on the maximum eigenvalue so it is reasonable that the
stability results for the dynamic generated by this matrix is
different from the one enunciated above. The matrix $\M$ also
contributes to the dynamic and, due to its particular form, gives
some unexpected result, namely if $c_I \not= c_E$ the average of
the contribution of $\M$ to the dynamic goes like $t\sqrt n$ so it
is divergent for large $n$ but it goes in any case to zero due to
the multiplication with  the exponential $e^{-\kappa t}$. Thus we
expect in this case to have large coherent motions which are then
dumped by the exponential factor. If $c_I=c_E$ the situation is
completely different because the term of the order $t\sqrt n$ is
multiplied by a constant equal to zero and disappears. In this
case the contribution of $\M$ converges to a gaussian random
variable with zero mean and variance of the type of a constant $+
J_0(\sqrt\sigma_* t)$ and so we get the usual stability result.
These are the meaning of the theorems demonstrated in the paper.

\section{Notations and formulations of the results}
Consider a dynamical system with random interactions (so-called a
complex system in \cite{[May]}) defined by
\begin{equation}
\x  '=-\kappa\x+{\J}'\x,
 \label{dyn}\end{equation}
where $\x\in\mathbb{R}^n$, $\kappa$ is a real number and $\J'$ is
an $n\times n$ real random matrix. Following Rajan and Abbott (see
(\cite{AR})) we consider the case, when $J_{ij}$ are independent
Gaussian variables with mean values
\begin{equation} E\{J_{ij}'\}=a\cdot\left\{\begin{array}{ll}
\mu_I/n^{1/2},&j=1,\dots [fn],\\
\mu_E/n^{1/2},& otherwise.\end{array} \right.
  \label{EJ}\end{equation}
Here $0<f<1$ and $a>0$ are fixed parameters, and $\mu_I$ and $\mu_E$
are chosen so  that the vector $\m=(m_1,\dots,m_n)$
\begin{equation} m_i=\left\{\begin{array}{ll}
\mu_I/n^{1/2},&j=1,\dots [fn]\\
\mu_E/n^{1/2}&, otherwise,\end{array} \right.
  \label{m}\end{equation}
 satisfies conditions
\begin{equation}\label{cond_mu}
    (\m,\u)=0,\quad (\m,\m)=1
\end{equation}
with
\begin{equation}\label{u}
   \u= (1,\dots,1).
\end{equation}
The variances of $J_{ij}$ are chosen as follows
\begin{equation} E\{(J_{ij}'-E\{J_{ij}'\})^2\}=n^{-1}\sigma_j=\left\{\begin{array}{ll}
\sigma_I/n,&j=1,\dots [fn],\\
\sigma_E/n,& otherwise.\end{array} \right.
  \label{DJ}\end{equation}
It is easy to see that in this case the matrix $J'$ could be
represented in the form
\begin{equation}\label{J+M}
   \J'=\J+a\M,
\end{equation}
where the matrix $\M$ is a rang  one matrix  all rows  equal to
$\m$, so that
\begin{equation}\label{M}
    \M\x=(\m,\x)\u, \quad \x\in\mathbb{R}^n,
\end{equation}
and the matrix $\J$ has the form
\begin{equation}\label{J=W}
    \J=n^{-1/2}\W
\end{equation}
where $\W$ a Gaussian matrix with independent entries satisfying
conditions
\begin{equation}\label{W}
    E\{W_{ij}\}=0,\quad E\{W_{ij}^2\}=\sigma_j
\end{equation}
with $\sigma_j$ defined in (\ref{DJ}).

 As it was shown numerically in the paper \cite{AR}, the matrix $\J'$ under
conditions (\ref{EJ})-(\ref{DJ}) has a spectrum which is not
localized in some fixed domain of $\mathbb{C}$ and so it is
difficult to expect that the dynamics of the system (\ref{dyn}) will
be stable. But if  we introduce the additional equilibrium
conditions
\begin{equation}\label{equi}
    \J \u=0\Leftrightarrow\sum_{j=1}^n
    W_{ij}=0\quad(i=1,\dots,n),
\end{equation}
then the spectrum $\J'$  coincides with the spectrum of $\J$, which
is well localized according to the results of \cite{Gir,Bai}.

It is easy to see, that under conditions (\ref{cond_mu}), (\ref{M})
and (\ref{equi})
\begin{equation}\label{MJ}
    \M^2=0,\quad \J \M=0
\end{equation}
so that for any $k\ge 1$
\[(\J+a\M)^k=
\J^k+a\M\J^{k-1}.\] Hence
\begin{equation}\label{e^J+M}
e^{t(\J+\M)}=e^{t\J}+a\int_0^tds\M e^{s\J}
\end{equation}
and  the solution of the system (\ref{dyn}) could be represented in
the form
\begin{equation}\label{sol}
x_i(t)=e^{-\kappa t}(e^{t\J}\x(0))_i+ae^{-\kappa
t}\int_0^tds(e^{s\J}\x(0),\m),
\end{equation}
where $\x(0)$ is a vector of initial conditions. Thus  to study
the dynamics (\ref{dyn}) it suffices to study the dynamics of the
system
\begin{equation}\label{dyn1}
\x'=\J \x
\end{equation}
with a matrix $\J$ of the form (\ref{J=W}), and   $\W$, satisfying
conditions (\ref{W}) and (\ref{equi}).

Supply the system with the initial conditions
\begin{equation}\label{in_c}
x_i(0)=c_i+\xi_i,
\end{equation}
where $\{\xi_i\}$ are independent random variables with distributions $\{\nu_i\}$, satisfying
the conditions
\begin{equation}\label{x0}
    \quad E\{\xi_i\}=0,\quad E\{(\xi_i)^2\}=\sigma^{(0)}_i,\quad E\{(\xi_i)^8\}\le C.
\end{equation}
and
\begin{equation}\label{c}
    c_i=\left\{\begin{array}{ll}
c_I,&i=1,\dots [fn],\\
c_E,& otherwise,\end{array} \right.\quad \nu_i(x)
=\left\{\begin{array}{ll}
\nu_I(x),&i=1,\dots [fn],\\
\nu_E(x),& otherwise,\end{array} \right. \quad \sigma^{(0)}_i
=\left\{\begin{array}{ll}
\sigma^{(0)}_I,&i=1,\dots [fn],\\
\sigma^{(0)}_E,& otherwise.\end{array} \right.
\end{equation}
Define the normalized counting function of $x_i$, solutions of the
system (\ref{dyn1}),
\begin{equation}
\N_n(\lambda,t)=n^{-1}\sharp\{x_i(t)\le\lambda\}=n^{-1}\sum_{i=1}^n\theta(\lambda-x_i(t)),
 \label{N}\end{equation}
where $\theta(x)$ is the standard Heaviside function.
$\N_n(\lambda,t)$ is a random measure on the real line which
counts the fraction of the variables $x_1,\dots,x_n$ which are
less then $\lambda$ at time $t$. Thus it characterizes the
distribution  of $x_i(t)$ on the real line.
 \begin{theorem}\label{thm:1}
 Consider the system (\ref{dyn1}) with a matrix $\J$ of the form (\ref{J=W}) under conditions
 (\ref{W})  and (\ref{equi}), and supply this system by the initial conditions (\ref{in_c})-(\ref{c}).
 Then for any $t>0$,
\begin{equation}
\lim_{n\to\infty}\N_n(\lambda,t)=f\N_I(\lambda,t)+(1-f)\N_E(\lambda,t)
 \label{a(t)}\end{equation}
where $\N_I(\lambda,t)$ and $\N_E(\lambda,t)$ are the convolutions
of the initial distribution $\nu_I$ and $\nu_E$ with normal distributions  $\N(c_I,\tilde\sigma(t))$
and $\N(c_E,\tilde\sigma(t)$ respectively
\begin{equation}
\N_I(\lambda,t)=\left(\nu_I*\N(c_I,\tilde\sigma(t))\right)(\lambda),\quad
\N_E(\lambda,t)=\left(\nu_E*\N(c_E,\tilde\sigma(t))\right)(\lambda),
 \label{N_I,E}\end{equation}
  and  the variance $\tilde\sigma(t)$ has the form
\begin{equation}
\tilde\sigma(t)=A\sigma_*^{-1}\sumd_{m=1}^\infty\frac{\sigma_*^m
t^{2m}}{m!m!}.
 \label{sigma}\end{equation}
 where
\begin{equation}\begin{array}{l}
\sigma_*=f\sigma_I+(1-f)\sigma_E,\\
A=\sigma_*^{-1}\sigma_I\sigma_Ef(1-f)(c_I-c_E)^2+
(\sigma_I\sigma_I^{(0)}f+\sigma_E\sigma_E^{(0)}(1-f)).
\end{array} \label{A}\end{equation}
 \end{theorem}

\medskip

\begin{theorem}\label{thm:2}
 Consider the system (\ref{dyn1}) with  matrix $\J$  of the form (\ref{J=W}) under conditions
 (\ref{W}) and (\ref{equi}), and supply this system by the initial conditions (\ref{in_c}) with (\ref{x0}).
 Set
\begin{equation}
w_n(t)=\int_0^tds(e^{s\J}\x(0),\m). \label{w(t)}\end{equation} If
$c_I\not=c_E$, then
\begin{equation}\label{}
    E\{w_n(t)\}=n^{1/2}t(fc_I\mu_I+(1-f)c_E\mu_E)\sim\sqrt n.
\end{equation}
If $c_I=c_E$, then  $w_n(t)$ for each fixed $t$ converges in
distribution to a Gaussian random variable with zero mean and
variance
\begin{equation}\label{}
\tilde\sigma^{(0)}=(1-f)\sigma^{(0)}_I+f\sigma^{(0)}_E+
A\sigma_*^{-1}\sum_{m=1}^\infty\frac{\sigma_*^mt ^{2m}}{m!m!},
\end{equation}
where $A$ and $\sigma_*$ are defined in (\ref{A}).
 \end{theorem}

\section{Proofs}

First of all we need to compute the expectations of $E\{W_{ij}\}$
and $E\{W_{ij}W_{kl}\}$ under  conditions (\ref{equi})
\begin{lemma}\label{lem:1}
(i) Under  conditions (\ref{equi}) and (\ref{W})
\begin{equation}\label{l1.1}
E\{W_{ij}\}=0,\quad
E\{W_{ij}W_{kl}\}=\delta_{ik}\left(\delta_{jl}\sigma_j-
\frac{\sigma_j\sigma_l}{n\sigma_*}\right),
\end{equation}
where
\begin{equation}\label{s_*}
\sigma_*=n^{-1}\sum\sigma_k=n^{-1}([fn]\sigma_I+(n-[fn])\sigma_E).
\end{equation}
(ii) Consider the random variable of the form
\begin{equation}\label{(W,d)}
  z=\sum_{k=1}^nn^{-1/2}W_{1k}d_k,
\end{equation}
 where the coefficients $d_k$ do not depend  on $\{W_{1k}\}$. Then $z$ is a normal variable
 with zero mean and the variance
 \begin{equation}\label{s_z}
    \sigma_z=n^{-1}\sum d_k^2\sigma_k-\sigma_*^{-1}(n^{-1}\sum d_k\sigma_k)^2
\end{equation}
\end{lemma}
{\bf Proof of Lemma \ref{lem:1}} The first equality in (\ref{l1.1}
is evident, because conditions (\ref{equi}) are symmetric with
respect to the change $W_{ik}\to -W_{ik}$. Besides, since different
lines of the matrix $\W$ have independent entries it is evident that
for $i\not=k$ $E\{W_{ij}W_{kl}\}=0$, and for $i=k$
$E\{W_{kj}W_{kl}\}$ do not depend on $k$. Hence,
\begin{multline}\label{l1.2}
E\{W_{kj}W_{kl}\}=E\{W_{1j}W_{1l}\}=\lim_{\e\to 0}\frac{\int
W_{1j}W_{1l}\exp\{-\sum W_{1j}^2/2\sigma_j -(\sum
W_{1j})^2/(2n\e)\}d\mathbf{W}}{\int \exp\{-\sum W_{1j}^2/2\sigma_j
-(\sum W_{1j})^2/(2n\e)\}d\mathbf{W}}\\
=\lim_{\e\to 0}\frac{\int dW_{1j}W_{1l}\exp\{-\sum
W_{1j}^2/2\sigma_j +it\sum n^{-1/2}W_{1j}-\e
t^2/2\}d\mathbf{W}dt}{\int \exp\{-\sum W_{1j}^2/2\sigma_j +it\sum
n^{-1/2}W_{1j}-\e t^2/2\}d\mathbf{W}dt}=
\end{multline}
$$ =\frac{\int
(\delta_{jl}\sigma_j-n^{-1}t^2\sigma_j\sigma_l)\exp\{-t^2\sum
\sigma_j/(2n) \}dt}{\int \exp\{-t^2\sum \sigma_j/(2n) \}dt}=$$
$$
=\delta_{jl}\sigma_j- \frac{\sigma_j\sigma_l}{n\sigma_*},$$

where $d\mathbf{W}=\prodd_{j=1}^ndW_{kj}$.

To prove the assertion (ii) of Lemma \ref{lem:1} we compute by the
same way the characteristic function of $z$
\begin{multline}\label{l1.3}
E\{e^{isz}\}=\lim_{\e\to 0}\frac{\int \exp\{-\sum
W_{1j}^2/2\sigma_j +is\sum_{k=1}^nn^{-1/2}W_{1k}d_k-(\sum
W_{1j})^2/(2n\e)\}d\mathbf{W}}{\int \exp\{-\sum W_{1j}^2/2\sigma_j
-(\sum W_{1j})^2/(2n\e)\}d\mathbf{W}}\\
=\lim_{\e\to 0}\frac{\int \exp\{-\sum
W_{1j}^2/2\sigma_j+is\sum_{k=1}^nn^{-1/2}W_{1k}d_k +it\sum
n^{-1/2}W_{1j}-\e t^2/2\}d\mathbf{W}dt}{\int \exp\{-\sum
W_{1j}^2/2\sigma_j +it\sum n^{-1/2}W_{1j}-\e
t^2/2\}d\mathbf{W}dt}=
\end{multline}
$$=\frac{\int \exp\{-\sum (t+sd_k)^2\sigma_j/(2n) \}dt}{\int
\exp\{-t^2\sum \sigma_j/(2n) \}dt}=e^{-s^2\sigma_z/2}.$$

Lemma \ref{lem:1} is proved.

\medskip

Below it will be convenient to consider the matrix $\J$ in the new
orthonormal basis. Denote
\begin{equation}\label{E_1,2}E_1=\hbox{Lin}\{\mathbf{e}_1,\dots,\mathbf{e}_{[fn]}\},\quad
E_2=\hbox{Lin}\{\mathbf{e}_{[fn]+1},\dots,\mathbf{e}_n\},
\end{equation}
where $\{\mathbf{e}_1,\dots,\mathbf{e}_{n}\}$ is the  basis in which
we consider the system (\ref{dyn1}) initially, so that
$x_i=(\x,\mathbf{e}_i)$.

Then define in $E_1$ and $E_2$ the orthonormal systems
$\{\u_3,\dots,\u_{[fn]+1}\}$ and $\{\u_{[fn]+2},\dots,\u_n\}$ which
are orthogonal to the vectors
$\mathbf{e}_1+\dots+\mathbf{e}_{[fn]}$, and
$\mathbf{e}_{[fn]+1}+\dots+\mathbf{e}_n$ respectively. If we denote
\begin{equation}\label{u_1,2}
\u_1=n^{-1/2}\u,\quad \u_2=m,
\end{equation}
then, according to (\ref{u}), (\ref{cond_mu}) and our choice of
$\u_3,\dots,\u_n$, the system $\{\u_i\}_{i=1}^n$ forms an
orthonormal basis in $\mathbb{R}^n$. Let $(u_{1i},\dots,u_{ni})$ be
the components of the vector $\u_i$ in the basis
$\{\mathbf{e}_1,\dots,\mathbf{e}_{n}\}$. Then the matrix
\begin{equation}\label{m_U}
\mathbf{U}=\{u_{ki}\}_{k,i=1}^n
\end{equation}
is a matrix of the orthogonal transformation from the basis
$\{\mathbf{e}_1,\dots,\mathbf{e}_{n}\}$ to the basis
$\{\u_1,\dots,\u_n\}$. Consider the matrix $\J$ in this basis.
\begin{equation}\label{ti_J}
    \tilde \J=\mathbf{U}^
    *\J\mathbf{U}=n^{-1/2}\mathbf{U}^*\W\mathbf{U}=n^{-1/2}\tilde\W.
\end{equation}
\begin{lemma}\label{lem:2}

The entries $\{\tilde W_{ki}\}_{k,i=1}^n$ are independent Gaussian
variables with zero means and their variances are
\begin{equation}\label{l2.1}
 E\{\tilde W_{ij}^2\}=\left\{\begin{array}{ll}
0,&j=1,\\
\sigma_I,&j=3,\dots [fn]+1,\\
\sigma_E,&j= [fn]+2\dots,n,\\
\sigma_I\sigma_E/\sigma_*,&j=2.
\end{array} \right.
\end{equation}
\end{lemma}
\begin{remark}\label{rem:1}
It follows from Lemma \ref{lem:2} that the matrix $\tilde\J$ can be
represented in the form
\[\tilde\J=\tilde\J_1\mathbf{D}^{(1)},\]
where $\tilde\J_1$ is a matrix with i.i.d. entries Gaussian entries
with zero means and variances $1$, and $\mathbf{D}^{(1)}$ is a diagonal
matrix with $D^{(1)}_{11}=0$,
$D^{(1)}_{22}=(\sigma_I\sigma_E/\sigma_*)^{1/2}$
 $D^{(1)}_{ii}=\sigma_I^{1/2}$ for $i=3,\dots [fn]+1$ and $D^{(1)}_{ii}=\sigma_E^{1/2}$ for
$i= [fn]+2\dots,n$.
 Hence
 \[||\tilde\J||^2\le \max\{\sigma_I,\sigma_E\}||\tilde\J_1||^2 .\]
  Using the result of \cite{Gir}, according to which under
condition    matrix $\tilde\J_1$ with i.i.d.
\[\P\{||\tilde\J_1||^2>4+\varepsilon\}\le e^{-C_1n\varepsilon^{2}},\]
and the fact that $||\tilde J||=||\J||$, we obtain now that
\begin{equation}
 \P\{||\J||>2L+\varepsilon\}\le e^{-Cn\varepsilon^{2}},
     \label{||J||}\end{equation}
 where we denote
  \begin{equation}
L=\max\{\sigma_I^{1/2}, \sigma_E^{1/2} \}.
       \label{L}\end{equation}
\end{remark}
\medskip

\begin{remark}\label{rem:2}
 Inequality (\ref{||J||}) allows us to use $||\J||$ in our considerations like a
bounded random variables. Indeed, since, e.g., $|x_1(t)|\le
ne^{t||\J||}$, denoting
$P_n(\lambda)=\emph{\P}\{||\J||>2L+\lambda\}$ and using
(\ref{||J||}), we can write for any fixed $t$ and $m,s<<n/\log n$
$$
E\{|x_1(t)|^me^{s||\J||}\}\le$$
$$\le
e^{s(2L+\epsilon)}E\{|x_1(t)|^m\theta(2L+\epsilon-||\J||)\}\\+
n^mE\{e^{(s+mt)||\J||}\theta(||\J||-2L-2\epsilon)\}\le$$
$$ \le e^{s(2L+\epsilon)}E\{|x_1(t)|^m\}
+n^m\int_{\lambda>\epsilon}e^{(s+mt)\lambda}dP_n(\lambda)\\
\le e^{s(2L+\epsilon)}E\{|x_1(t)|^m\}+O(e^{-Cn\varepsilon^2/2}).
$$
Hence, below we use $||\J||$ as a bounded variable without
additional explanations.
\end{remark}
\medskip

{\bf Proof of Lemma \ref{lem:2}}. It is evident that $\{\tilde
W_{ki}\}_{k,i=1}^n$ have joint Gaussian distribution, so to prove
Lemma \ref{lem:2} it is enough to compute their covariances. To this
aim we use relation
\begin{equation}\label{l2.2}
\tilde W_{ij}=\sum_{k,l} u_{ki}W_{kl}u_{lj}
\end{equation}
and Lemma \ref{lem:1}. Then from the first equality of \ref{l1.1} we
derive that the mean values of $\{\tilde W_{ki}\}_{k,i=1}^n$ are
equal to zero.
\begin{equation}\label{l2.3}
E\{\tilde W_{i_1j_1}\tilde W_{i_2j_2}\}=\sum
u_{k_1i_1}u_{l_1j_1}u_{k_2i_2}u_{l_2j_2} E\{W_{k_1l_1}W_{k_2l_2}\}.
\end{equation}
Now we use the fact that for different $k_1$ and $k_2$ $W_{k_1l_1}$
$W_{k_2l_2}$  are independent and for $k_1=k_2$
$E\{W_{k_1l_1}W_{k_1l_2}\}$ does not depend on $k_1$ (see
(\ref{l1.1})). Substituting \ref{l1.1}) in (\ref{l2.3}),  summing
with respect to $k_1$ and using the orthogonality of $\u_{i_1}$ and
$\u_{i_2}$, we get
\begin{equation}\label{l2.5}
E\{\tilde W_{i_1j_1}\tilde W_{i_2j_2}\}=\delta_{i_1i_2}\sum
u_{l_1j_1}u_{l_2j_2}(\delta_{l_1l_2}
\sigma_{l_1}-\sigma_{l_1}\sigma_{l_2}/(\sigma_{*}n)).
\end{equation}
Now if $j_1=1$, then $u_{l_1j_1}=n^{-1/2}$ and summation with
respect to $l_1$ gives us zero because of (\ref{s_*}). If
$j_1=3,\dots,[fn]+1$, then, since  $u_{l_1j_1}=0$ for
$l_1\ge[fn]+1$, we have that in the r.h.s. of (\ref{l2.5})
$\sigma_{l_1}=\sigma_I$, and so, using the orthogonality of
$\u_{j_1}$ and $\u_{j_2}$, we get the second line of (\ref{l2.1}).
If $j_1=[fn]+2,\dots,n$ the proof is the same. Now we are left to
prove the last line of (\ref{l2.1}). Using (\ref{l2.5}) and
(\ref{cond_mu}), which gives us
\begin{equation}\label{mu_1,2}
\mu_I=-\sqrt{(n-[fn])/[fn]},\quad \mu_E=\sqrt{[fn]/(n-[fn])},
\end{equation}
we obtain
\begin{multline}\label{l2.6}
E\{\tilde W_{i2}\tilde W_{i2}\}=n^{-1}\sum
_{l_1=1}^{[fn]}\sigma_I\mu_I^2+ n^{-1}\sum
_{l_1=[fn]+1}^{n}\sigma_E\mu_E^2 -(\sigma_*n^2)^{-1}\left(\sum
_{l_1=1}^{[fn]}\sigma_I\mu_I+
\sum_{[fn]+1}^n\sigma_E\mu_E\right)^2=
\end{multline}
$$
=(\sigma_I(1-[fn]/n)+\sigma_E[fn]/n)-\frac{(\sigma_I-\sigma_E)^2[fn](n-[fn])}{\sigma_*n^2}=$$
$$ =\frac{\sigma_I\sigma_E}{\sigma_*}.$$

{\bf Proof of Theorem \ref{thm:1}} Let us consider the system
(\ref{dyn1})  from the second equation  to the last one as a system
of equations for $x_2(1),\dots,x_n(t)$, where $x_1(t)$ is a known
function. Then
\begin{equation}
x_i(t)=x^{(1)}_i+\sum_{j=2}^n\int_0^tds
(e^{(t-s)\J^{(1)}})_{ij}\frac{W_{j1}}{n^{1/2}}x_1(s),
 \label{t1.2}\end{equation}
where
\begin{equation}
x^{(1)}_i=(e^{t\J^{(1)}}\x(0))_i
 \label{t1.3}\end{equation}
 and $\J^{(1)}$ is the matrix which we obtain from $\J$ replacing the first
line and the first column by zeros. Substituting this expressions in
the first equation of (\ref{dyn1}), we get
\begin{equation}
x_1'(t)=\sum_{j=2}^n\frac{W_{1j}}{n^{1/2}}x_j^{(1)}(t)+\int_0^tds
\,\tilde r_n^{(1)}(t-s)x_1(s)+\frac{W_{11}}{n^{1/2}}x_1(t),
 \label{t1.4}\end{equation}
where
\begin{equation}
\tilde
r_n^{(1)}(t)=n^{-1}\sum_{i,j=2}^n(e^{t\J^{(1)}})_{ij}W_{1i}W_{j1}.
 \label{t1.5}\end{equation}
Hence
\begin{equation}
x_1(t)=x_1(0)+\sum_{j=2}^n\frac{W_{1j}}{n^{1/2}}d_j^{(1)}(t)+\int_0^tds \,
r_n^{(1)}(t-s)x_1(s),
 \label{t1.6}\end{equation}
with
\[
d_j^{(1)}(t)=\int_0^tds \,x_j^{(1)}(s),\quad r_n^{(1)}(t)=\int_0^t
d\tau\, \tilde r_n^{(1)}(\tau)+n^{-1/2}W_{11}.
\]
Using Lemma \ref{lem:1}, it is easy to see that
\begin{equation}
E\{(\tilde r_n^{(1)}(t))^8\}\le C(t)n^{-4}.
 \label{t1.7}\end{equation}
 Indeed, according to (\ref{t1.5}),
 \[\tilde r_n^{(1)}=n^{-1/2}\sum_{i=2}^nW_{1i}f_i,\quad f_i=
 n^{-1/2}\sum_{j=2}^n(e^{t\J^{(1)}})_{ij}W_{j1}\]
 with $f_i$ independent of $\{W_{1j}\}$. Hence,
 if we take the eighth  power of (\ref{t1.5}) and take the expectation with respect to $\{W_{1j}\}$,
 we get
 \begin{equation}\label{t1.7a}
 E\{(\tilde r_n^{(1)}(t))^8\}=
 \end{equation}
 $$=7\cdot 5\cdot 3\cdot  E\left\{\left(n^{-2}\sum_{j_1,j_2}(e^{t\J^{(1)T}}
 \mathbf{D}e^{t\J^{(1)}})_{j_1j_2}W_{j_11}W_{j_21}-
 \right.\right.\\
 -\left.\left.n^{-3}\sigma_*^{-1}(\sum_{i_1,j_1}\sigma_i(e^{t\J^{(1)}})_{i_1j_1}W_{j_11})^2\right)^4\right\} \le$$
$$ \le
105n^{-4}E\left\{||D||e^{8t||J^{(1)}||}\left(n^{-1}\sum_{j_1}W^2_{j_11}\right)^2\right\}\le
$$
$$ \le C(t)n^{-4},$$

where $\J^{(1)T}$ means the transposed matrix of $\J^{(1)T}$, $\mathbf{D}$ is a diagonal matrix such that
\begin{equation}
D_{ij}=\delta_{ij}\sigma_i\label{D},
\end{equation}
and here and below we denote by $C(t)$ function of $t$ (different
in different formulas), such that
\[C(t)\le Ce^{ct}\]
with some $n$-independent $C$ and $c$.

The relation  (\ref{t1.7a}) and a trivial crude bound
\[|x_1(t)|\le ||e^{t\J}\x(0)||\le e^{t||\J||}||x(0)||\le C(t)\sqrt n\]
allow us to obtain
\begin{equation}
E\{x_1^4(t)\}\le
\end{equation}
$$
\le 27\left(E\{x_1^4(0)\}+3L^2 \left(n^{-1}\sum(
d_i^{(1)}(t))^2\right)^2\right.\\\left.+t^3\int_0^t ds
E^{1/2}\{(r_n^{(1)}(t-s))^8\} E^{1/2}\{x_1^8(s)\}\right) \le$$
$$
\le 27 \left(C+3L^2 E\{(n^{-1}||e^{t\J^{(1)}}\x(0)||^2)\}+
C(t)n^{-4}\int_0^tds
E^{1/2}\{x_1^4(s)e^{4s||\J||}||x(0)||^4\}\right)\le $$
$$ \le
C(t)\left(1+n^{-2}\int_0^tdsE^{1/2}\{x_1^4(s)\}\right),$$

 where $L$ is defined by (\ref{L}).
 Then, by a standard argument, we get
\begin{equation}\label{x^4}
E\{x_1^4(t)\}\le C(t).
\end{equation}
This bound allows us to write (\ref{t1.6}) as
\begin{equation}\label{t1.8}
x_1(t)=x_1(0)+\sum_{j=2}^n\frac{W_{1j}}{n^{1/2}}d_j^{(1)}(t)+\e_n^{(1)}(t),
\end{equation}
where
\begin{equation}\label{t1.9}
E\{(\e_n^{(1)}(t))^2\}\le C(t)n^{-1}.
\end{equation}
Now we can apply Lemma \ref{lem:1}, which gives us that the sum in
the r.h.s. of (\ref{t1.8}) is a normal random variable with the
variance
\begin{equation}\label{t1.10}
\tilde\sigma_n^{(1)}(t)=n^{-1}\sum
\sigma_j(d_j^{(1)}(t))^2-\sigma_*^{-1}(n^{-1}\sum\sigma_id^{(1)}_i)^2.
\end{equation}
Define
\begin{equation}\label{R_n}
R_n(t,s)= n^{-1}\sum_{j=1}^n \sigma_jx_j(t)x_j(s).
\end{equation}
 \begin{lemma}\label{lem:3}
 Under conditions of Theorem \ref{thm:1} $R_n(t,s)$
 \begin{equation}\label{D_n}
 D_n(t,s)=E\{(R_n(t,s)-E\{R_n(t,s)\})^2\} \le C(t)C(s)n^{-1},
\end{equation}
\begin{equation}\label{s-a.1}
 E\bigg\{\bigg(n^{-1}\sum_{i=1}^n\sigma_ix_i(t)-
  E\bigg\{n^{-1}\sum_{i=1}^n\sigma_ix_i(t)\bigg\}\bigg)^2\bigg\} \le C(t)n^{-1}.
\end{equation}
Besides,
 \begin{equation}\label{l3.1}
E\{ |R_n(t,s)-n^{-1}\sum_{j=1}^n
\sigma_jx_j^{(1)}(t)x_j^{(1)}(s)|^2\}\le C(t)C(s)n^{-1}.
 \end{equation}
\end{lemma}
Denote
\begin{equation}\label{t1.10a}
\tilde\sigma_n(t)=\int_0^t\int_0^tdt'ds'R_n(t',s')
-\sigma_*^{-1}(\int_0^tdt'n^{-1}\sum\x_i(t'))^2.
\end{equation}
\begin{remark}\label{rem:3}
Lemma \ref{lem:3} implies
\begin{equation}\label{t1.10b}
E\{(\tilde\sigma_n^{(1)}(t)-\tilde\sigma_n(t))^2\}\le C(t)n^{-1},
\end{equation}
\begin{equation}\label{t1.10c}
E\{(\tilde\sigma_n(t)-E\{\tilde\sigma_n(t)\})^2\}\le C(t)n^{-1}.
\end{equation}
\end{remark}
{\bf Proof of Lemma \ref{lem:3}}. To prove (\ref{l3.1}) we first
estimate
\begin{equation}\label{l3.2}
n^{-1}\sum_{j=1}^n \sigma_j(x_j(t)-x_j^{(1)}(t))(x_j(s)-x_j^{(1)}(s))=\\
\end{equation}
$$
=n^{-1}\sum_{j_1,j_2=2}^n\int_0^sds_1 \int_0^tds_2
 (e^{(t-s_1)\J^{(1)T}}\mathbf{D}e^{(s-s_2)\J^{(1)}})_{j_1j_2}\frac{W_{1j_1}W_{1j_1}}{n}x_1(s_1)x_1(s_2)
 \le $$
 $$
\le n^{-2}tse^{(s+t)||\J^{(1)}||}\left(\int_0^t
x_1^2(s_1)ds_1\right)^{1/2}
 \left(\int_0^s x_1^2(s_2)ds_2\right)^{1/2}\sum_{j=2}^nW_{j1 }^2.
 $$

 Now, using the Schwartz inequality  we get
 \begin{multline}\label{l3.3}
 \bigg|R_n(t,s)-n^{-1}\sum_{j=2}^n \sigma_jx_j^{(1)}(t)x_j^{(1)}(s)\bigg|^2=
 \bigg|n^{-1}\sum_{j=1}^n \sigma_j\left((x_j(t)-x_j^{(1)}(t))x_j(s)+x_j^{(1)}(t)(x_j(s)-x_j^{(1)}(s))\right) \bigg|^2
 \\ \le \left(n^{-1}\sum_{j=1}^n \sigma_j\left((x_j(t)-x_j^{(1)}(t))^2+(x_j(s)-x_j^{(1)}(s))^2\right)\right)
  \left(n^{-1}\sum_{j=1}^n \sigma_j\left(x_j^2(s)+(x_j^{(1)}(t))^2\right)\right) \\
 \le Cn^{-3}tse^{2(s+t)||\J^{(1)}||}\left(\int_0^t x_1^2(s_1)ds_1+\int_0^s x_1^2(s_2)ds_2\right)||x(0)||^2
 \sum_{j=2}^nW_{j1 }^2.
 \end{multline}
    Combining this with (\ref{x^4}) we obtain (\ref{l3.1}).
To prove (\ref{D_n}), we write
\begin{equation}\label{l3.4}
 D_n(t,s)=\sum_{i,j=1}^n\sigma_i\sigma_j
 E\bigg\{\bigg(x_i(t)x_i(s)-E\{x_i(t)x_i(s)\}\bigg)\bigg(x_j(t)x_j(s)-E\{x_j(t)x_j(s)\}\bigg)\bigg\}.
\end{equation}
Let us estimate, e.g.
$E\bigg\{\bigg(x_1(t)x_1(s)-E\{x_1(t)x_1(s)\}\bigg)\bigg(x_2(t)x_2(s)-E\{x_2(t)x_2(s)\}\bigg)\bigg\}$.
To this end we write the representations ( cf.(\ref{t1.6}))
\begin{equation}\label{l3.5}\begin{array}{rcl}
x_1(t)&=&x_1(0)+n^{-1/2}W_{12}d_2(t)+n^{-1/2}\sumd_{j=3}^nW_{1j}d_j^{(1,2)}(t)\\
&&+\intd_0^tdt'
\, r_n^{(1,1)}(t-t')x_1(t')+\int_0^tdt' \,
r_n^{(1,2)}(t-t')x_2(t'),\\
x_2(t)&=&x_2(0)+n^{-1/2}W_{21}d_1(t)+n^{-1/2}\sumd_{j=3}^nW_{2j}d_j^{(1,2)}(t)\\
&&+\intd_0^tdt'
\, r_n^{(2,1)}(t-t')x_1(t')+\int_0^tdt' \, r_n^{(2,2)}(t-t')x_2(t'),
 \end{array}\end{equation}
where
\[
d_j^{(1,2)}(t)=\int_0^tds \,x_j^{(1,2)}(s),\quad x_j^{(1,2)}(t)=(e^{t\J^{(1,2)}}\x(0))_{j},\quad
r_n^{(\alpha,\beta)}(t)=\int_0^t d\tau\, \tilde
r_n^{(\alpha,\beta)}(\tau),
\]
\[
\tilde
r_n^{(\alpha,\beta)}(t)=n^{-1}\sum_{i,j=2}^n(e^{t\J^{(1,2)}})_{ij}W_{\alpha
i}W_{j\beta}.
\]
 $\J^{(1,2)}$ is the matrix $\J$ with the first and the second lines and the first and the second
 columns replaced by zeros. Similarly to (\ref{t1.7})-(\ref{t1.9}) we obtain
\begin{equation}\label{l3.7}\begin{array}{rcl}
x_1(t)&=&x_1(0)+n^{-1/2}\sumd_{j=3}^nW_{1j}d_j^{(1,2)}(t)
\, +\e_n^{(1)}(t),\\
x_2(t)&=&x_2(0)+n^{-1/2}\sumd_{j=3}^nW_{2j}d_j^{(1,2)}(t)+\e_n^{(2)}(t),\\
&&E\{(\e_n^{(1)}(t))\}\le n^{-1}C(t),\,\,E\{(r_n^{(2)}(t))\}\le n^{-1}C(t).
 \end{array} \end{equation}
Then we can write
\begin{multline}\label{l3.8}
E\bigg\{\bigg(x_1(t)x_1(s)-E\{x_1(t)x_1(s)\}\bigg)\bigg(x_2(t)x_2(s)-E\{x_2(t)x_2(s)\}\bigg)\bigg\}\\
\le
\int_0^t\int_0^s\int_0^t\int_0^sdt_1'ds_1'dt_2'ds_2'
E\bigg\{\bigg(R^{(1,2)}(t_1',s_1')-E\{R^{(1,2)}(t_1',s_1')\}\bigg)\\
\bigg(R^{(1,2)}(t_2',s_2')-E\{R^{(1,2)}(t_2',s_2')\}\bigg)\bigg\}+C(t)C(s)n^{-1}\\
\le\int_0^t\int_0^s\int_0^t\int_0^sdt_1'ds_1'dt_2'ds_2'\left(
E\bigg\{\bigg(R^{(1,2)}(t_1',s_1')-E\{R^{(1,2)}(t_1',s_1')\}\bigg)^2\bigg\}\right.\\+
\left.E\bigg\{\bigg(R^{(1,2)}(t_2',s_2')-E\{R^{(1,2)}(t_2',s_2')\}\bigg)^2\bigg\}\right)+C(t)C(s)n^{-1}\\
\le 2ts\int_0^t\int_0^sdt'ds'E\bigg\{\bigg(R^{(1,2)}(t',s')-E\{R^{(1,2)}(t',s')\}\bigg)^2\bigg\}
+C(t)C(s)n^{-1},
 \end{multline}
where
\[R^{(1,2)}(t,s)=n^{-1}\sum_{i=3}^n\sigma_ix_i^{(1,2)}(t)x_i^{(1,2)}(s).
\]
Similarly to (\ref{l3.1})
\[E\{(R_n^{(1,2)}(t,s)-R_n(t,s))^2\}\le C(t)C(s)n^{-1}.
\]
Repeating (\ref{l3.8}) for all terms in (\ref{l3.4}) with different $i,j$,
we obtain the inequality
\[
D_n(t,s)\le 2\sigma_*^2ts\int_0^t\int_0^sdt'ds'D_n(t',s').
\]
Iterating this inequality $M=[\log n]$ times, we get (\ref{D_n}).
The proof of the inequality (\ref{s-a.1}) follows from the representation (\ref{l3.5})
immediately, if we use the independence of $x_1(0)$ and $x_2(0)$.

Lemma \ref{lem:3} is proved.

\medskip

Now we are ready to prove the self averaging property of $\N_n(\lambda,t)$, as
 $n\to\infty$, i.e. we prove that for any real $\lambda$
and $t>0$
\begin{equation}
\lim_{n\to\infty}E\bigg\{\bigg(\N_n(\lambda,t)-E\{\N_n(\lambda,t)\}\bigg)^2\bigg\}=0.
 \label{self-av_N}\end{equation}
According to the standard theory of measure, for this aim it is
enough to prove that  $g_n(z,t)$ -- the Stieltjes
transform of the distribution $\N_n(\lambda,t)$
\begin{equation}
g_n(z,t)=\int\frac{d\N_n(\lambda,t)}{\lambda-z}=n^{-1}\sum_{i=1}^n\frac{1}{x_i(t)-z},\quad(\Im
z\not=0 ),
 \label{g_n}\end{equation}
 for any $z:\Im z\not=0$ possesses  a self averaging.
 property.
\begin{lemma}\label{lem:4} For any $z:\Im z\not=0$
\begin{equation}
\lim_{n\to\infty}E\bigg\{\bigg|g_n(z,t)-E\{g_n(z,t)\}\bigg|^2\bigg\}=0.
 \label{self-av}\end{equation}
\end{lemma}
{\bf Proof of Lemma \ref{lem:4}.}
Similarly to (\ref{l3.4}) we write
\begin{multline}
E\bigg\{\bigg|g_n(z,t)-E\{g_n(z,t)\}\bigg|^2\bigg\}\\=
n^{-2}\sum_{i,j=1}^n\bigg(E\bigg\{(x_i(t)-z)^{-1}(x_j(t)-\overline z)^{-1}\bigg\}-
E\bigg\{(x_i(t)-z)^{-1}\bigg\}E\bigg\{(x_j(t)-\overline z)^{-1}\bigg\}\bigg).
 \label{l4.1}\end{multline}
Then repeating the arguments (\ref{l3.4})-(\ref{l3.7}), we obtain
\begin{multline}
E\bigg\{\bigg|g_n(z,t)-E\{g_n(z,t)\}\bigg|^2\bigg\}\\
\le
C(t)n^{-1}+n^{-2}\sum_{i,j=1}E\bigg\{\bigg(F_i(\tilde\sigma_n,z)-E\{F_i(\tilde\sigma_n,z)\}\bigg)
\bigg(F_j(\tilde\sigma_n,z)-E\{F_j(\tilde\sigma_n,z)\}\bigg)\bigg\},
 \label{l4.2}\end{multline}
where $\tilde\sigma_n$ is defined by (\ref{t1.10a}) and we denote
\[
F_i(\sigma,z)=\frac{1}{\sqrt{2\pi}}\int\int \frac{d\nu_i(x)dye^{-y^2/2}}{x+c_i+\sigma^{1/2}y-z}.
\]
Since evidently
\[|F_i(\sigma_1,z)-F_i(\sigma_2,z)|\le C\frac{|\sigma_1^{1/2}-\sigma_2^{1/2}|}{|\Im z|^2},\]
we get from (\ref{l4.2})
\begin{equation}\label{l4.3}
E\bigg\{\bigg|g_n(z,t)-E\{g_n(z,t)\}\bigg|^2\bigg\}\le C(t)n^{-1}+C|\Im z|^{-2}
E\{(\tilde\sigma_n-E\{\tilde\sigma_n\})^2\}.
\end{equation}
Now the assertion of Lemma \ref{lem:4} follows from (\ref{t1.10c}).

\medskip

Using  (\ref{l3.1}), the s.a. properties (\ref{D_n}),  and the fact that the system (\ref{dyn1})
is symmetric with respect to $x_1,\dots,x_{[fn]}$ and with respect
$x_{[fn]+1},\dots, x_n$, we obtain
\begin{multline}\label{t1.11}
E\{\tilde\sigma_n(t)\}= \frac{\sigma_I[fn]}{n}E\left\{\left(\int_0^tds
x_1(s)\right)^2\right\}+ \sigma_E(1-[fn]/n)E\left\{\left(\int_0^tds
x_n(s)\right)^2\right\}
-\\
\sigma_*^{-1}\left(\frac{\sigma_I[fn]}{n}\int_0^tds E\{x_1(s)\}+
\sigma_E(1-[fn]/n)\int_0^tds E\{x_1(s)\}\right)^2+o(1).
\end{multline}
Repeating our conclusions for $x_n(t)$, we get
\begin{equation}\label{t1.12}
x_n(t)=x_n(0)+\sum_{j=2}^n\frac{W_{nj}}{n^{1/2}}d_j^{(n)}(t)+\e_n^{(n)}(t),
\end{equation}
where
\begin{equation}\label{t1.13}
E\{(\e_n^{(n)}(t))^2\}\le C(t)n^{-1},
\end{equation}
\[d_i^{(n)}(t)=\int_0^tx_i^{(n)}(s)ds\quad(i=1,\dots,n-1),
\]
 $x_i^{(n)}=(e^{t\J^{(n)}}\x(0))_i$
and the matrix $\J^{(n)}$ is obtained from $\J$ by replacing the
last line and the last column by zeros. Then, applying Lemma
\ref{lem:1},  we obtain that the second sum in (\ref{t1.12}) is a
Gaussian random variable with the same variance $\tilde\sigma_n(t)$
(see (\ref{t1.11})).

Equations (\ref{t1.10}) and (\ref{t1.12}) combined with (\ref{in_c})
give us
\begin{equation}\label{t1.14}
E\{x_1(t)\}=c_I,\quad E\{x_n(t)\}=c_E.
\end{equation}
Denoting
\begin{equation}\label{ER}
R^{(1)}_n(t,s)=E\{x_1(t)x_1(s)\},\quad R^{(2)}_n(t,s)=E\{x_n(t)x_n(s)\},
\end{equation}
we obtain from (\ref{t1.10}) and (\ref{t1.12}) the system of
equations
\begin{equation}\label{RR}\begin{array}{lcl}
R^{(1)}_n(t,s)&=&E\{x_1^2(0)\}+\sigma_If\intd_0^t\int_0^sR^{(1)}_n(t',s')dt'ds'\\
&&+\sigma_E(1-f)\intd_0^t\int_0^sR^{(2)}_n(t',s')dt'ds'
-ts\sigma_*^{-1}(\sigma_Ifc_1+
\sigma_E(1-f)c_2)^2+o(1),\\
R^{(2)}_n(t,s)&=&E\{x_n^2(0)\}+\sigma_If\intd_0^t\int_0^sR^{(1)}_n(t',s')dt'ds'\\
&&+\sigma_E(1-f)\intd_0^t\int_0^sR^{(2)}_n(t',s')dt'ds'
-ts\sigma_*^{-1}(\sigma_Ifc_1+ \sigma_E(1-f)c_2)^2+o(1).
\end{array}\end{equation}
Then we obtain that the function
\begin{equation}\label{R^0}
R^{(0)}_n(t,s)=\sigma_IfR^{(1)}_n(t,s)+\sigma_E(1-f)R^{(2)}_n(t,s)-\sigma_*^{-1}(\sigma_Ifc_1+
\sigma_E(1-f)c_2)^2,
\end{equation}
satisfies the equation
\begin{equation}\label{eq_R}
R^{(0)}_n(t,s)=A+\sigma_*\int_0^t\int_0^sR^{(0)}_n(t',s')dt'ds'+o(1),
\end{equation}
where
\begin{equation}
A=\sigma_IfE\{x_1^2(0)\}+\sigma_E(1-f)E\{x_n^2(0)\}-\sigma_*^{-1}(\sigma_Ifc_1+
\sigma_E(1-f)c_2)^2=
\end{equation}
$$=
\frac{\sigma_I\sigma_Ef(1-f)(c_1-c_2)^2}{\sigma_*}+
(\sigma_I\sigma_I^{(0)}f+\sigma_E\sigma_E^{(0)}(1-f)).$$

As it was proved in \cite{FST1} this equation has the unique
solution
\begin{equation}
R^{(0)}_n(t,s)=A\left(1+\sum_{m=1}^\infty
\frac{\sigma_*^mt^ms^m}{m!m!}\right) +o(1).
\label{exp_R}\end{equation}

Then we can easily find that
\begin{equation}\label{RR1}\begin{array}{lcl}
R^{(1)}_n(t,s)&=&E\{x_1^2(0)\}+A\sigma_*^{-1}\sumd_{m=1}^\infty
\frac{\sigma_*^mt^ms^m}{m!m!}+o(1),\\
R^{(2)}_n(t,s)&=&E\{x_n^2(0)\}+A\sigma_*^{-1}\sumd_{m=1}^\infty
\frac{\sigma_*^mt^ms^m}{m!m!}+o(1).
\end{array}\end{equation}
Hence
\begin{equation}\label{ti-s}
\tilde\sigma(t)=\lim_{n\to\infty}
\intd_0^t\int_0^tdt'ds'R^{(0)}_n(t',s')=A\sigma_*\sigma_*^{-1}\sumd_{m=1}^\infty
\frac{\sigma_*^mt^{2m}}{m!m!}.
\end{equation}
Using (\ref{self-av_N}), and the symmetry of the problem
we obtain that $\mathcal{N}_n(t,\lambda)$ converges in probability to
\[fE\{\theta(\lambda-x_1(t))\}+(1-f)E\{\theta(\lambda-x_n(t))\}.\]
But by the above arguments
\[\begin{array}{l}
E\{\theta(\lambda-x_1(t))\}\to\intd_{-\infty}^\lambda\frac{dy}{\sqrt{2\pi}}\int d\nu_I(x)
\exp\{(x-\tilde\sigma^{1/2}(t)y-c_I)^2/2\},\\
E\{\theta(\lambda-x_n(t))\}\to\intd_{-\infty}^\lambda \frac{dy}{\sqrt{2\pi}}\int d\nu_E(x)
\exp\{(x-\tilde\sigma^{1/2}(t)y-c_E)^2/2\}.
\end{array}\]
Theorem 1 follows.

\medskip

{\bf Proof of Theorem \ref{thm:2}}. To prove Theorem \ref{thm:2} it
is convenient to consider the system (\ref{dyn1}) in the basis
$\{\mathbf{u}_i\}_{i=1}^n$ defined above (see
(\ref{E_1,2})-(\ref{m_U}). Let
\begin{equation}\label{y}
y_i(t)=(\x(t),\mathbf{u}_i)
\end{equation}
Then the system (\ref{dyn1}) takes the form
\begin{equation}\label{dyn2}
\mathbf{y}'=\tilde \J\mathbf{y},
\end{equation}
where $\tilde \J$ is defined by (\ref{ti_J}). The question of
interest is the behavior of $y_2(t)$. Repeating for $y_2(t)$ the
arguments (\ref{t1.2})-(\ref{t1.7}) we get the representation
\begin{equation}
y_2(t)=y_2(0)+\tilde J_{22}\int_0^ty_2(s)ds+\sum_{j=3}^n\frac{\tilde
W_{1j}}{n^{1/2}}d_j(t)+\int_0^tds \, r_n^{(2)}(t-s)y_2(s),
 \label{t2.2}\end{equation}
with
\begin{equation}\label{y(0)}
y_2(0)=(x(0),m)=\sqrt n
(c_I\mu_If+c_E\mu_E(1-f))+\mu_I\sum_{i=1}^{[fn]}\frac{x^{(0)}_i}{\sqrt
n}+ \mu_E\sum_{i=[fn]+1}^n\frac{x^{(0)}_i}{\sqrt n},
\end{equation}
\[\begin{array}{l}
d_j(t)=\intd_0^tds \,y_j^{(2)}(s),\quad y^{(2)}_i=(e^{t\tilde\J^{(2)}}\mathbf{y}(0))_i,\\
r_n^{(2)}(t)=\intd_0^t \tilde r_n^{(2)}(\tau)d\tau+n^{-1/2}W_{22},\quad \tilde
r_n^{(2)}(t)=n^{-1}\sumd_{i,j=3}^n(e^{t\tilde\J^{(2)}})_{ij} \tilde
W_{2i}\tilde W_{j2},
\end{array}\]
where  and $\tilde \J^{(2)}$ is the matrix which we obtain from
$\tilde\J$ replacing the first and the second lines and the first
and the second columns by zeros. Taking the expectation in
(\ref{t2.2}) we get the first statement of Theorem \ref{thm:2}. Now
assume that $c_I=c_E=c$. Then, repeating arguments
(\ref{t1.5})-(\ref{x^4}), we get that
\begin{equation}
y_2(t)=y_2(0)+\sum_{j=3}^n\frac{\tilde W_{1j}}{n^{1/2}}d_j(t)+
\tilde\e_n(t),
 \label{t2.3}\end{equation}
 and
\begin{equation}
E\{(\tilde \e_n^{(2)}(t))^2\}\le C(t)n^{-1}.
 \label{t2.3}\end{equation}
Applying Central Limit Theorem to the r.h.s. of (\ref{y(0)}) it is
easy to obtain that $y_2(0)$ converges in distribution to a Gaussian
random variable with zero mean and the variance
\[
\sigma_y=f\mu_I^2\sigma_I+(1-f)\mu_E^2\sigma_E=(1-f)\sigma_I+f\sigma_E.
\]
Besides, since $\{\tilde W_{2,j}\}$ are independent Gaussian random
variables, the sum in the r.h.s. of (\ref{t2.2}) is a gaussian
random variable with the variance
\begin{multline}
\tilde\sigma^{(1)}(t)=n^{-1}\sigma_I\sum_{j=3}^{[fn]+1}(d^{(2)}_j(t))^2+
n^{-1}\sigma_E\sum_{j=[fn]+2}^{n}(d^{(2)}_j(t))^2=\\=
n^{-1}\sigma_I\sum_{j=3}^{[fn]+1}\int_0^t\int_0^tdt'ds'y_j(t')y_j(s')+
n^{-1}\sigma_E\sum_{j=[fn]+2}^{n}\int_0^t\int_0^tdt'ds'y_j(t')y_j(s')+o(1)=\\
\label{t2.4}\end{multline}
$$
=n^{-1}\sigma_I\int_0^t\int_0^tdt'ds'\sum_{k,k'=1}^nx_k(t')x_{k'}(s')\sum_{j=3}^{[fn]+1}u_{kj}u_{k'j}+
n^{-1}\sigma_E\int_0^t\int_0^tdt'ds'\sum_{k,k'=1}^nx_k(t')x_{k'}(s')\sum_{j=[fn]+2}^nu_{kj}u_{k'j}.$$

If  $1\le k\le [fn]$, while $[fn]+1\le k'\le n$ or $1\le k'\le
[fn]$, while $[fn]+1\le k\le n$, then by construction of
$\mathbf{u}_3,\dots,\mathbf{u}_n$
\[\sum_{j=3}^{[fn]+1}u_{kj}u_{k'j}=0,\quad \sum_{j=[fn]+2}^nu_{kj}u_{k'j}=0.\]
Let $1\le k,k'\le [fn]$. Then
\[\sum_{j=[fn]+2}^nu_{kj}u_{k'j}=0\]
and so
\[\sum_{j=3}^{[fn]+1}u_{kj}u_{k'j}=
\sum_{j=3}^{n}u_{kj}u_{k'j}=\delta_{k,k'}-u_{k1}u_{k'1}-u_{k2}u_{k'2}=
\delta_{k,k'}-n^{-1}(1+\mu_I^2).\] Similarly for $[fn]+1\le k,k'\le
n$, we get
\[
\sum_{j=[fn]+2}^nu_{kj}u_{k'j}=\delta_{k,k'}-n^{-1}(1+\mu_E^2).
\]
Finally we get
\begin{equation}
\tilde\sigma^{(1)}(t) =\label{t2.4}\end{equation}
$$
=n^{-1}\sigma_I\int_0^t\int_0^tdt'ds'\sum_{k,k'=1}^{[fn]}x_k(t')x_{k'}(s')(\delta_{k,k'}-
n^{-1}(1+\mu_I^2))+$$
$$+
n^{-1}\sigma_E\int_0^t\int_0^tdt'ds'\sum_{k,k'=[fn]+1}^nx_k(t')x_{k'}(s')(\delta_{k,k'}-
n^{-1}(1+\mu_E^2)=$$
$$
=f\sigma_I\int_0^t\int_0^tdt'ds'R^{(1)}(t',s')+(1-f)\sigma_E\int_0^t\int_0^tdt'ds'R^{(2)}(t',s')
-t^2(\sigma_If^2c^2(1+\mu_I^2)+\sigma_E(1-f)^2c^2(1+\mu_E^2))=$$
$$
=\int_0^t\int_0^tdt'ds'R^{(0)}(t',s')=A\sigma_*^{-1}\sum_{m=1}^\infty+o(1).
\frac{\sigma_*^mt^ms^m}{m!m!}$$

Since the sum of independent Gaussian variables is a Gaussian random
variable with the variance equal to the sum of variances, we obtain
that $w_n(t)$ converge in distribution to a Gaussian random variable
with zero mean and the variance
\[\tilde\sigma^{(0)}=\tilde\sigma^{(1)}(t)+\sigma_y.\]

The second assertion of Theorem \ref{thm:2} follows.

\end{document}